\documentclass[12pt]{article}

\usepackage{amsmath,amssymb,epsfig}
\usepackage{graphicx}
\usepackage{booktabs}

\begin{document}
\setlength{\parskip}{0.45cm}
\setlength{\baselineskip}{0.75cm}

%
%
%
\begin{titlepage}
\setlength{\parskip}{0.25cm}
\setlength{\baselineskip}{0.25cm}
\begin{flushright}
DO-TH 06/02\\
\vspace{0.2cm}
April 2006
\end{flushright}
\vspace{1.0cm}
\begin{center}
\Large
{\bf Non--singlet QCD analysis of $F_2(x,Q^2)$ up to NNLO}
\vspace{1.5cm}

\large
M.~Gl\"uck, E.~Reya, C.~Schuck\\
\vspace{1.0cm}

\normalsize
{\it Institut f\"{u}r Physik´, Universit\"{a}t Dortmund}\\
{\it D-44221 Dortmund, Germany} \\
\vspace{0.5cm}

\vspace{1.5cm}
\end{center}

\begin{abstract}
\noindent The significance of NNLO (3--loop) QCD contributions to the 
flavor non--singlet sector of $F_2^{ep}$ and $F_2^{ed}$ has been studied
as compared to uncertainties (different factorization schemes, higher twist
and QED contributions) of standard NLO (and LO) QCD analyses. The latter
effects turn out to be comparable in size to the NNLO contributions.
Therefore the minute NNLO effects are unobservable with presently
available (precision) data on non--singlet structure functions.
\end{abstract}
\end{titlepage}


\section{Introduction}
In a recent publication \cite{ref1} a next--to--next--to--leading order
(NNLO) QCD analysis of $F_2^{ep}(x,Q^2)$ and $F_2^{ed}(x,Q^2)$ in the 
flavor non--singlet sector was presented.  Here our purpose is to
study the significance of the NNLO contribution as compared to other, 
possibly important, contributions such as redundant terms in the NLO
analysis (which arise when the NLO evolved parton distributions are multiplied
by the coefficient function), higher twist and QED contributions to NLO,
and effects due to choosing different factorization schemes.  In Sec.~2
we present the relevant theoretical expressions required for our analysis,
and Sec.~3 contains the quantitative results.  Our conclusions are summarized
in Sec.~4.

\section{Theoretical Formalism}
The non--singlet (NS) parts of the structure functions $F_2^{ep,d}(x,Q^2)$
for $x>0.3$, where valence quark dominance is adopted, are, at LO,
given by
\begin{eqnarray}
  F_2^{ep} = \frac{4}{9}\,  x\, u_v+\frac{1}{9}\, x\, d_v & \equiv &
    \frac{5}{18}\, x\, q_{{\rm NS,}8}^+ + \frac{1}{6}\, x\, q_{{\rm NS,}3}^+\\
  F_2^{ed} = \frac{5}{18}\, x(u_v+d_v) & \equiv & 
    \frac{5}{18}\, x\, q_{{\rm NS,}8}^+
\end{eqnarray}
where  
$d=(p+n)/2$ and $q_{{\rm NS,}3}^+=u_v-d_v$.  For $x<0.3$ one analyzes the 
genuine NS combination
\begin{equation}
F_2^{p-n}\equiv 2 (F_2^{ep}-F_2^{ed}) = \frac{1}{3}\, x(u_v-d_v) 
  +\frac{2}{3}\, x(\bar{u}-\bar{d}) \equiv \frac{1}{3}\, x\, q_{{\rm NS,}3}^+
\end{equation}
where now $q_{{\rm NS,}3}^+ = u_v-d_v+2(\bar{u}-\bar{d})$ since sea quarks 
cannot be neglected for $x$ smaller than about 0.3.  For definiteness
we adopt for $\bar{d}-\bar{u}$ the choice \cite{ref2,ref1}
\begin{equation}
x(\bar{d}-\bar{u})(x,Q_0^2) = 1.195 x^{1.24}(1-x)^{9.1}
  (1 + 14.05x - 45.52 x^2)
\end{equation}
at $Q_0^2=4$ GeV$^2$ which gives a good description of the Drell--Yan
dimuon production data \cite{ref3}, but plays a marginal role in our
analysis.

At NLO$(\overline{\rm MS})$ the $n$--th Mellin moments of the above
NS combinations of valence parton distributions, for brevity denoted
by $v^+$, symbolically evolve according to the well known expression
(see, e.g.\ \cite{ref4,ref5})
\begin{equation}
v^+(Q^2)=\left \{ 1-(a-a_0)R_1\right \}\, 
  (a/a_0)^{-P_{{\rm NS}}^{(0)}/\beta_0}\,  v^+ (Q_0^2)
\end{equation}
where $R_1=P_{{\rm NS}}^{(1)+}/\beta_0 - (\beta_1/\beta_0^2)P_{{\rm NS}}^{(0)}$
and $a=a(Q^2)\equiv \alpha_s(Q^2)/4\pi$ with $a_0=a(Q_0^2)$.
The moments of the above NS structure functions $F_2^{{\rm NS}}$ are 
then given by
\begin{equation}
 F_2^{{\rm NS}}(Q^2) = \left[ 1+aC_{2,{\rm NS}}^{(1)}\right] v^+(Q^2)
\end{equation}
and this expression is commonly compared with experiment.  Inserting 
$v^+(Q^2)$ from (5) into this equation one observes a {\em redundant}
${\cal{O}}(a^2)$ contribution, i.e.\ $-a(a-a_0)C_{2,{\rm NS}}^{(1)}R_1$,
which in fact belongs to a NNLO analysis and is assumed to be small 
at NLO.  If, however, one chooses to work to a NNLO (3--loop) accuracy,
such a redundancy at NLO might become significant as compared to the 
full NNLO contribution.  This will be investigated quantitatively below.
Similarly, the choice of a factorization scheme, other than the 
$\overline{\rm MS}$ scheme used thus far, might imply larger differences
than additional NNLO contributions in the $\overline{\rm MS}$ scheme.
For example, in the deep inelastic scattering (DIS) factorization 
scheme \cite{ref6,ref4} the Wilson coefficient in (6) is absorbed into
the parton distributions, i.e.\ into their evolutions in (5) with
$R_1\to R_1^{\rm DIS} = R_1-C_{2,{\rm NS}}^{(1)}$~:
\begin{equation}
v_{\rm DIS}^+(Q^2) =\left\{ 1 -(a-a_0)R_1^{\rm DIS}\right\}
   (a/a_0)^{-P_{\rm NS}^{(0)}/\beta_0}\,  v^+(Q_0^2)
\end{equation}
and, instead of (6), $F_2^{\rm NS} = v_{\rm DIS}^+$ at NLO.

Taking into acccount QCD 3--loop NNLO $\alpha_s^3$ effects, one also has
to consider QED LO $\alpha$--contributions which are of comparable size
\cite{ref7}.  The latter can easily be implemented by changing the 
$n$--th moments of the NLO valence input distribution $q_v(Q_0^2)$,
$q_v=u_v,\, d_v$, in (5) according to
\begin{equation}
q_v(Q_0^2)\to (a/a_0)^{\frac{\alpha}{4\pi}\, \frac{\beta_1}{\beta_0^2}P^{\gamma}}
\exp \left[\frac{\alpha}{4\pi\beta_0}\left(\frac{1}{a}-\frac{1}{a_0}\right)
 P^{\gamma}\right]\, q_v(Q_0^2)
\end{equation}
with $\alpha\simeq \frac{1}{137}$ and $P^{\gamma}=(e_q^2/C_F)P_{\rm NS}^{(0)}$
where $C_F=4/3$.

At NNLO($\overline{\rm MS}$) the evolution of $v^+(Q^2)$ in (5) generalizes
to 
\begin{equation}
v^+(Q^2) = \{ 1-(a-a_0)R_1-\frac{1}{2}(a^2-a_0^2)(R_2-R_1^2)
 -a_0(a-a_0)R_1^2\}(a/a_0)^{-P_{\rm NS}^{(0)}/\beta_0}\, v^+(Q_0^2)
\end{equation}
with $R_2=P_{\rm NS}^{(2)+}/\beta_0-(\beta_1/\beta_0)R_1-
   (\beta_2/\beta_0^2)P_{\rm NS}^{(0)}$ and (6) becomes
\begin{equation}
F_2^{\rm NS}(Q^2) 
  = \left[1+aC_{2,{\rm NS}}^{(1)}+a^2C_{2,{\rm NS}}^{(2)+}\right]\, 
          v^+(Q^2)\, .
\end{equation}
Convenient expressions for the relevant 2--loop Wilson coefficient
$C_{2,{\rm NS}}^{(2)+}$ can be found in \cite{ref5} (eq.~(A.2)) and
for the 3--loop splitting function $P_{\rm NS}^{(2)+}$ in \cite{ref8}
(eq.~(4.22), which can be easily Mellin--transformed using 
\cite{ref9}). 
The strong coupling now evolves according to 
$da/d\ln Q^2 =-\sum_{\ell =0}^2 \beta_{\ell}\, a^{\ell +2}$ 
where 
$\beta_0 = 11-2f/3$, $\beta_1 = 102-38f/3$ 
and $\beta_2 = 2857/2-5033f/18+325f^2/54$,
which refers to the $\overline{\rm MS}$ renormalization scheme, and
$f$ denotes the number of active flavors.  Here the redundant 
${\cal{O}}(a^3,a^4)$ contributions in (10) turn out to be marginal and
do not influence our (fit) results.  
Furthermore, the running coupling $a(Q^2)$ is appropriately matched 
at $Q= m_b = 4.5$ GeV and $Q= m_t = 175$ GeV. To obtain $F_2^{\rm NS}$
in the DIS factorization scheme at NNLO, the Wilson coefficient 
functions in (10) have to be absorbed into the parton distributions,
i.e. into their evolutions in (9) with 
$R_2\to R_2^{\rm DIS}=R_2-2C_{2,\rm NS}^{(2)+}+( C_{2,\rm NS}^{(1)})^2$:
\begin{eqnarray}
v_{\rm DIS}^+ (Q^2) & = & \{ 1-(a-a_0)R_1^{\rm DIS} -\frac{1}{2}(a^2-a_0^2)\,
  \left[ R_2^{\rm DIS}-(R_1^{\rm DIS})^2 \right]  \nonumber\\
& & -a_0(a-a_0)(R_1^{\rm DIS})^2 \}\, (a/a_0)^{-P_{\rm NS}^{(0)}/\beta_0}
     \, v^+(Q_0^2)
\end{eqnarray}
where $R_1^{\rm DIS}$ is as in the NLO-DIS expression (7) and, instead
of (10), we now have \mbox{ $F_2^{\rm NS}=v_{\rm DIS}^+$ } 
at NNLO.
 
Since flavor NS structure functions are mainly related to the medium
and large $x$--region, the relevant kinematic nucleon target mass (TM)
corrections are always taken into account according to \cite{ref10}
\begin{eqnarray}
F_{2,{\rm TM}}^{\rm NS}(n,Q^2) 
& \equiv & 
  \int_0^1 x^{n-2} F_{2,{\rm TM}}^{\rm NS}(x,Q^2)\, dx = \nonumber\\
& = &
  \sum_{j=0}^2 \left(\frac{m_N^2}{Q^2}\right)^j\,
  \frac{(n+j)!}{j!(n-2)!}\,\,
  \frac{F_2^{\rm NS}(n+2j,Q^2)}{(n+2j)(n+2j-1)} 
+ {\cal{O}}\left( \left( \frac{m_N^2}{Q^2}\right)^3\right)
\end{eqnarray}
where higher powers than $(m_{\rm N}^2/Q^2)^2$ are negligible for the 
relevant $x <0.8$ region, as can straightforwardly be shown by comparing
(12) with the well known exact expression in Bjorken-$x$ space \cite{ref10}.

Despite the kinematic cuts ($Q^2\geq 4$ GeV$^2$,
$\, W^2\equiv (\frac{1}{x}-1)\, Q^2+m_{\rm N}^2\geq 10$ GeV$^2$) used
for our analysis, we also take
into account higher twist (HT) corrections to $F_2^{\rm NS}$ via
\begin{equation}
v^+(x,Q^2)\to \left[1+\frac{m_N^2}{Q^2}\,h(x)\right]\, v^+(x,Q^2)
\end{equation}
in order to learn whether nonperturbative effects may still contaminate
our perturbative analysis.  Here we adopt the ansatz \cite{ref11}
\begin{equation}
h(x) = a\left( \frac{x^b}{1-x} -c\right)\, .
\end{equation}

Notice that the input valence parton distributions $v^+(x,Q_0^2)$ at 
LO $[v^+(Q^2)=(a/a_0)^{-P_{\rm NS}^{(0)}/\beta_0}\,  v^+(Q_0^2)]$,
NLO and NNLO in the $\overline{\rm MS}$ as well as in the DIS scheme
in eqs.~(5)--(11) and (13) can and will be {\em different} in general. 

Finally, Fermi motion and nuclear effects in the deuteron are strongly
model dependent and will therefore not be considered here.  They were,
however, taken into account in \cite{ref1,ref11,ref12} using the specific
models cited there.  Comparing these results with our valence
distributions obtained and to be discussed below, as well as with
other results where such effects have not been taken into account
(e.g.~\cite{ref2}), shows that these effects do not change the 
quality of the QCD fits.

\section{Quantitative results}

In the present analysis we used the proton and deuteron data of BCDMS
\cite{ref13}, NMC \cite{ref14} and SLAC \cite{ref15}, as well as the 
proton data of H1 \cite{ref16} and ZEUS \cite{ref17} in the relevant
$x$--regions discussed above which amount to 480 data points.  The
valence distributions have been parametrized at the input scale
$Q_0^2=4$ GeV$^2$ as
\begin{eqnarray}
x\,  u_v(x,Q_0^2) & = & N_u\,  x^{a_u}(1-x)^{b_u}(1+A_u\,  
                   x^{c_u}+B_u\,  x)\\
x\,  d_v(x,Q_0^2) & = & N_d\,  x^{a_d}(1-x)^{b_d}(1+A_d\,  
                   x^{c_d}+B_d\,  x)
\end{eqnarray}
with the normalizations $N_u$ and $N_d$ being fixed by 
$\int_0^1 u_v dx=2$ and $\int_0^1 d_v dx=1$, respectively.
The LO, NLO and NNLO fit results without HT contributions
are summarized in Table 1.  The standard NLO and NNLO fits refer to
the $\overline{\rm MS}$ scheme according to eqs.~(5), (6) and (9), (10),
respectively.  Our fit results for NNLO are compared in 
Fig.~1 with the data used.  Except perhaps in LO,
we obtained equally good and acceptable fits ($\chi^2/$dof) in each
perturbative order and scenario.  As has been already noted previously
\cite{ref1,ref12,ref18,ref19}, a NNLO analysis in general results in a
slightly smaller $\alpha_s(m_Z^2)$ than in NLO.  
This is due to the fact that the higher the perturbative order the
faster $\alpha_s(Q^2)$ increases as $Q^2$ decreases.  In order to
compensate for this increase, a NNLO fit is expected to result in 
a smaller value for $\alpha_s(m_Z^2)$ than a NLO fit.
Notice that the values of $\alpha_s(m_Z^2)$ obtained in the usual
perturbative NLO and NNLO fits in Table 1 are comparable to the 
ones in \cite{ref1,ref12,ref18}. Repeating the NLO and NNLO fits in 
the DIS factorization scheme improves only marginally the global
$\overline{\rm MS}$ fits ($\chi^2$), and the QED ${\cal{O}}(\alpha)$ contributions
leave the original NLO($\overline{\rm MS}$) results practically 
unchanged as evident from Table 1.

On the other hand, the inclusion of higher twist contributions 
sizeably improves the fits, i.e., the value of $\chi^2/$dof as can be 
seen from Table 2 ($\alpha_s(m_Z^2)$ is reduced as expected since the 
HT term takes care already of some of the $Q^2$--dependence of the data). 
In order to illustrate the relative significance of the (model dependent)
HT corrections in (13), we have performed a fit for $Q^2\geq 4$ GeV$^2$
and one for $Q^2\geq 10$ GeV$^2$, denoted by HT(10) in Table 2.  
Clearly, HT effects
become less important when the lower cut of $Q^2$ is increased and the
value of $\chi^2$ increases, eventually approaching the larger values
obtained by purely perturbative fits.  Nevertheless it is remarkable
that the fits for $Q^2\geq 4$ GeV$^2$ and $Q^2\geq 10$ GeV$^2$ are not
significantly different as will be illustrated below (Fig.\ 2).
For illustration we also show in Fig.\ 1 the NNLO results with HT
effects included for $Q^2\geq 4$ GeV$^2$.  The results for the
$Q^2\geq 10$ GeV$^2$ cut are not shown, since they are very similar 
to the ones for the $Q^2\geq 4$ GeV$^2$ cut (dashed curves). 

The actual relative size of our results can best be seen by
comparing the various fit results with the pure QCD NLO fit, i.e.\
by considering the following ratios, depicted in Fig.~2, which are
defined as follows.  The effect of NNLO contributions can be visualized
via $r_{\rm NNLO}=F_{2,\rm NNLO}^{ep}/F_{2,\rm NLO}^{ep}$ 
with the nominal NLO structure function given by (6) and the NNLO
one by (10).  Similarly, $r_{\rm LO}$ requires the usual 
$F_{2,\rm LO}^{ep}$ as given by (5) and (6) with $R_1\equiv 0$ 
and  $C_2^{(1)}\equiv 0$.  Furthermore $r_{\rm NLO}^{\rm DIS}
=F_{2,\rm NLO}^{ep,\rm DIS}/F_{2,\rm NLO}^{ep}$
illustrates the effects of choosing the DIS factorization scheme
instead of the $\overline{\rm MS}$ scheme, with the DIS structure
function given by (7).  
Similarly, the definition of $r_{\rm NNLO}^{\rm DIS}$ employs the DIS
structure function in NNLO given by (11).
Including the QED ${\cal{O}}(\alpha)$ 
contributions to $F_{2,\rm NLO}^{ep}$ according to (8) results
in $r_{\rm NLO}^{\rm QED}=F_{2,\rm NLO}^{ep,\rm QED}/
F_{2,\rm NLO}^{ep}$. Figure 2(a) demonstrates that ambiguities of 
standard NLO $\overline{\rm MS}$ analyses or additional QED 
contributions are comparable in size to NNLO contributions in 
the relevant medium and large $x$--region.  
Only in the smaller $x$--region around $x= 0.2$, the NNLO results are
about 1\% larger than the NLO ones.  This difference, however, disappears
in the DIS factorization scheme where $r_{\rm NNLO}^{\rm DIS}$ and
$r_{\rm NLO}^{\rm DIS}$ are comparable and small (about 0.5\%). We
therefore conclude that the DIS scheme guarantees a better perturbative
convergence than the commonly used $\overline{\rm MS}$ scheme, except
in the very large $x$--region (where nonperturbative contributions
are uncontrollable anyway).  Unfortunately such minute effects are
not testable with presently available precision data for non--singlet
structure functions which have a typical uncertainty of about 10\% in
the small and large $x$--region.

The redundant ${\cal{O}}(a^2)$
contribution to a NLO analysis, as discussed after (6), turns out to
be marginal: repeating the NLO fit with the redundant term removed
from $F_{2,\rm NLO}^{ep}$ \mbox {in (6)},
\begin{equation}
F_{2,\rm NLO}^{ep,\rm rem} = F_{2,\rm NLO}^{ep} +a(a-a_0)
  C_{2,\rm NS}^{(1)}\, R_1(a/a_0)^{-P_{\rm NS}^{(0)}/\beta_0}\,
     v^+(Q_0^2)\, \, ,
\end{equation}
one obtains $|r_{\rm NLO}^{\rm rem}-1|=|F_{2,\rm NLO}^{ep,\rm rem}/
F_{2,\rm NLO}^{ep}-1|$ 
\raisebox{-0.1cm}{$\stackrel{<}{\sim}$} 0.001 
for all relevant values of $Q^2$.  

Finally, the relevance of HT 
effects is illustrated in Fig.~2(b) where the ratios 
$r_i^{\rm HT} = F_{2,i}^{ep,\rm HT}/F_{2,\rm NLO}^{ep}$
for the different perturbative orders are shown, with the HT
corrections incorporated according to (13).  
In general, for each perturbative order the fit results are {\it{stable}}
with respect to different choices for the lower bound on $Q^2$ in (13),
i.e., $Q^2\geq 4$ GeV$^2$ and 10 GeV$^2$ denoted by HT and HT(10),
respectively.  In particular the more relevant NLO and NNLO fit results
are very similar.  Despite the fact that these results are about twice
as large as the ones without HT contributions in Fig.\ 2(a) throughout
the whole $x$--region considered, they are still much smaller than
present experimental uncertainties.

\section{Conclusions}

The significance of NNLO QCD contributions to the flavor non--singlet
sector of $F_2^{ep,d}(x,Q^2)$ has been studied as compared to 
uncertainties of standard NLO (and LO) analyses.  NNLO corrections
slightly improve the fits to presently available data and imply a better
perturbative convergence in the DIS factorization scheme than in the
commonly used $\overline{\rm MS}$ scheme.
However, 
ambiguities of NLO fits such as the choice of a particular factorization
scheme ($\overline{\rm MS}$ vs.\ DIS) and possible higher twist effects
as well as QED ${\cal{O}}(\alpha)$ contributions turn out to be
comparable in size to NNLO (3--loop) contributions.  In particular,
nonperturbative higher twist effects play an important role in
obtaining optimal fits (minimal $\chi^2$) which turn out to be rather
stable with respect to different choices of the lower bound on $Q^2$.
Their contribution is
about twice as large as purely perturbative uncertainties which are
typically less than about 1\%.
We therefore
conclude that the rather minute NNLO QCD effects in the flavor
non--singlet sector are not observable with present precision data
for flavor non--singlet structure functions which have sizeably
larger errors.
\vspace{1.5cm}

\noindent This work has been supported in part by the  `Bundesministerium 
f\"ur Bildung und Forschung', Berlin/Bonn.



\clearpage
\pagestyle{empty}
\begin{table}
\centering
\footnotesize
\caption{Parameter values of the QCD fits in various perturbative 
orders based on all non--singlet data for $F_2^{ep,d}(x,Q^2)$. The 
parameters of the input valence distributions refer to (15) and (16). 
The analysis for the DIS factorization scheme is based on (7) for NLO
and on (11) for NNLO. The 
QED contribution to NLO is taken into account according to (8).}
\bigskip
\begin{tabular}{lrrrrc}
\toprule
&\multicolumn{3}{c}{{
Parameter}}
&${\chi^2}/{\rm{dof}}$
&$\alpha_s(m^2_Z)$\\
&$a_u$&$b_u$&$c_u$&&$\left({\Lambda^{(4)}}{/\mathrm{MeV}}\right)$\\
&$A_u$&$B_u$&$N_u$\\
&$a_d$&$b_d$&$c_d$\\
&$A_d$&$B_d$&$N_d$\\
\midrule
LO&$  0.574$&$  3.290$&$  0.823$&$0.98$&$ 0.128$\\
&$   9.258$&$  -7.164$&$   1.790$&&($196.0$)\\
&$   0.600$&$   4.952$&$   0.100$\\
&$  -0.330$&$   3.726$&$   1.742$\\[1.9mm]
NLO&$  0.600$&$  3.364$&$  0.667$&$0.93$&$ 0.112$\\
&$   9.900$&$  -8.504$&$   1.521$&&($222.6$)\\
&$   0.600$&$   5.163$&$   0.181$\\
&$   3.294$&$   8.303$&$   0.542$\\[1.9mm]
NLO DIS&$  0.600$&$  3.004$&$  0.855$&$0.91$&$ 0.113$\\
&$   9.801$&$  -9.010$&$   2.101$&&($228.3$)\\
&$   0.581$&$   4.797$&$   0.878$\\
&$  -0.030$&$   4.458$&$   1.264$\\[1.9mm]
NLO QED&$  0.600$&$  3.361$&$  0.667$&$0.93$&$ 0.112$\\
&$   9.883$&$  -8.496$&$   1.523$&&($217.1$)\\
&$   0.599$&$   5.161$&$   0.186$\\
&$   3.223$&$   8.037$&$   0.555$\\[1.9mm]
NNLO&$  0.600$&$  3.571$&$  0.599$&$0.89$&$ 0.111$\\
&$   9.654$&$  -6.810$&$   1.309$&&($177.2$)\\
&$   0.600$&$   5.209$&$   0.323$\\
&$   4.060$&$   4.870$&$   0.657$\\[1.9mm]
NNLO DIS&$  0.587$&$  2.727$&$  0.825$&$0.89$&$ 0.112$\\
&$   9.644$&$  -9.471$&$   1.941$&&($187.2$)\\
&$   0.600$&$   4.787$&$   0.868$\\
&$  -2.004$&$   6.046$&$   1.421$\\[1.9mm]
\bottomrule
\end{tabular}
\end{table}

\clearpage

\pagestyle{empty}
\begin{table}
\centering
\footnotesize
\caption{As in Table 1 but including HT contributions as well according
to (13) with the parameters ($a$, $b$, $c$) referring to (14). Fits
using a lower bound $Q^2\geq 10$ GeV$^2$ in (13) are denoted by
HT(10), whereas HT refers to $Q^2\geq 4$ GeV$^2$ as stated before (13).
NLO and NNLO always refer to the $\overline{\rm MS}$ factorization
scheme.}
\bigskip
\begin{tabular}{lrrrrc}
\toprule
&\multicolumn{3}{c}{{
Parameter}}
&${\chi^2}/{\rm{dof}}$
&$\alpha_s(m^2_Z)$\\
&$a_u$&$b_u$&$c_u$&&$\left({\Lambda^{(4)}}{/\mathrm{MeV}}\right)$\\
&$A_u$&$B_u$&$N_u$\\
&$a_d$&$b_d$&$c_d$\\
&$A_d$&$B_d$&$N_d$\\
&$a$&$b$&$c$\\
\midrule
LO HT&$  0.600$&$  3.218$&$  0.397$&$0.82$&$ 0.118$\\
&$   6.926$&$  -4.208$&$   1.021$&&($121.8$)\\
&$   0.582$&$   4.882$&$   0.712$\\
&$   7.205$&$  -2.878$&$   0.956$\\
&$   2.180$&$   0.941$&$   0.577$\\[1.9mm]
LO HT (10)&$  0.600$&$  3.167$&$  0.422$&$0.88$&$ 0.126$\\
&$   8.286$&$  -5.214$&$   0.942$&&($176.6$)\\
&$   0.597$&$   4.985$&$   0.750$\\
&$   6.163$&$  -1.061$&$   1.052$\\
&$   2.105$&$   1.000$&$   0.749$\\[1.9mm]
NLO HT&$  0.600$&$  3.368$&$  0.508$&$0.83$&$ 0.104$\\
&$   9.897$&$  -8.215$&$   1.104$&&($136.6$)\\
&$   0.600$&$   5.383$&$   0.222$\\
&$   4.905$&$   9.563$&$   0.467$\\
&$   2.198$&$   1.468$&$   0.303$\\[1.9mm]
NLO HT (10)&$  0.600$&$  3.429$&$  0.522$&$0.90$&$ 0.106$\\
&$   9.900$&$  -7.786$&$   1.126$&&($155.8$)\\
&$   0.598$&$   5.507$&$   0.190$\\
&$   2.383$&$   9.824$&$   0.642$\\
&$   2.313$&$   1.740$&$   0.284$\\[1.9mm]
NNLO HT&$  0.600$&$  3.571$&$  0.519$&$0.83$&$ 0.103$\\
&$   9.900$&$  -7.086$&$   1.116$&&($109.5$)\\
&$   0.566$&$   5.425$&$   0.302$\\
&$   5.816$&$   9.475$&$   0.441$\\
&$   1.695$&$   1.104$&$   0.432$\\[1.9mm]
NNLO HT (10)&$  0.600$&$  3.622$&$  0.518$&$0.91$&$ 0.105$\\
&$   9.898$&$  -6.491$&$   1.099$&&($124.9$)\\
&$   0.599$&$   5.594$&$   0.235$\\
&$   2.260$&$   9.119$&$   0.716$\\
&$   1.989$&$   1.801$&$   0.252$\\[1.9mm]
\bottomrule
\end{tabular}
\end{table}

\clearpage

\begin{figure}
\centering%
\includegraphics[height=\textwidth, angle=-90]{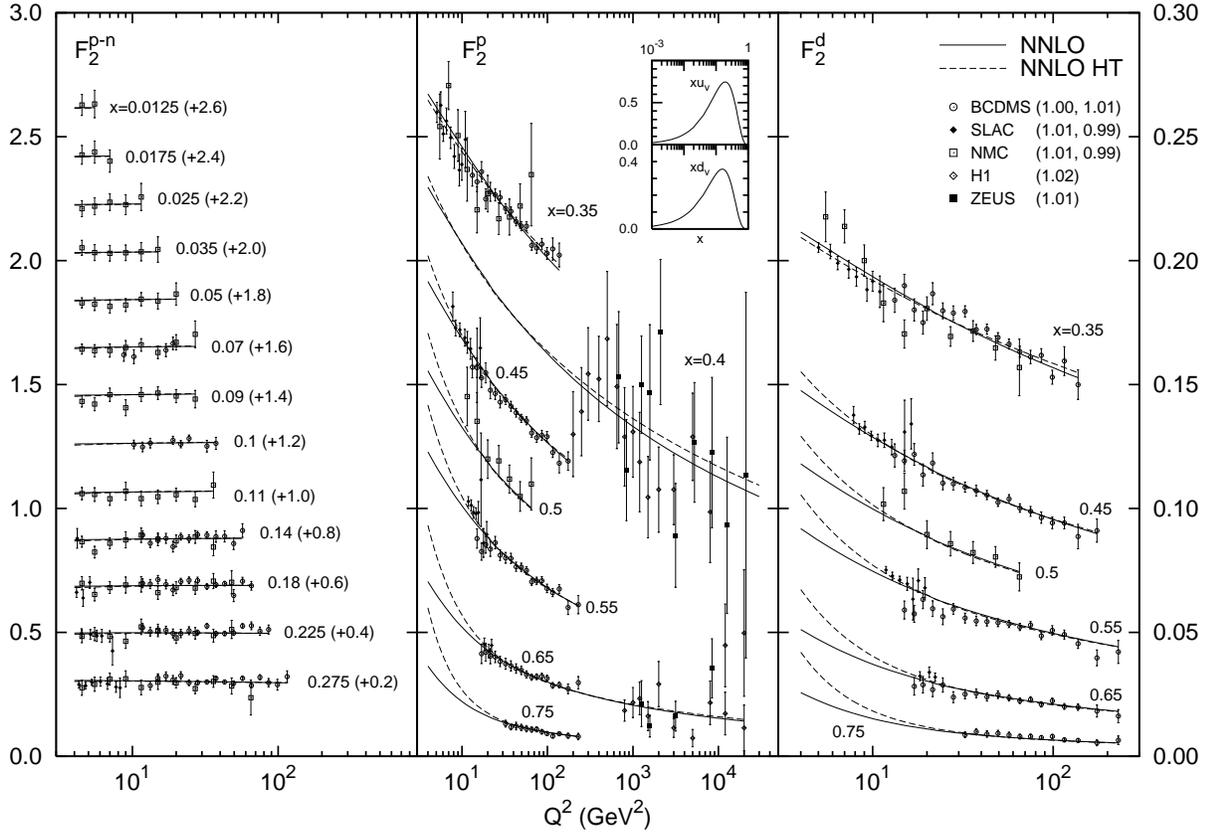}
\footnotesize
\caption{Comparison of our NNLO fits with all presently available 
flavor non--singlet data \cite{ref13,ref14,ref15,ref16,ref17} used 
for our analysis.
The higher twist (HT) contribution is taken into account according 
to (13) and (14). The NLO fits are very similar and 
practically indistinguishable from the ones shown. So is the NNLO HT(10)
fit resulting from the cut $Q^2\geq 10$ GeV$^2$. The inset shows 
our NNLO input valence distributions at $Q^2_0=4\,\mathrm{GeV}^2$. 
The scales on the left ordinate refer only to $F_2^{p-n}$ where for 
each fixed value of $x$ we have added the constant in brackets to 
$F_2^{p-n}$. The scales on the right ordinate refer to $F_2^p$ and 
to $F_2^d$. The data sets are shown with their normalization factors 
in parentheses (first entry refers to $F_2^p$, second entry to 
$F_2^d$) as obtained in the fit. The ZEUS data for $F_2^p$ have been 
shifted to the right by $5\,\%$ in order to make their error bars 
distinguishable from the ones of the H1 data.}
\end{figure}

\begin{figure}
\centering
\footnotesize
\includegraphics[width=0.7\textwidth]{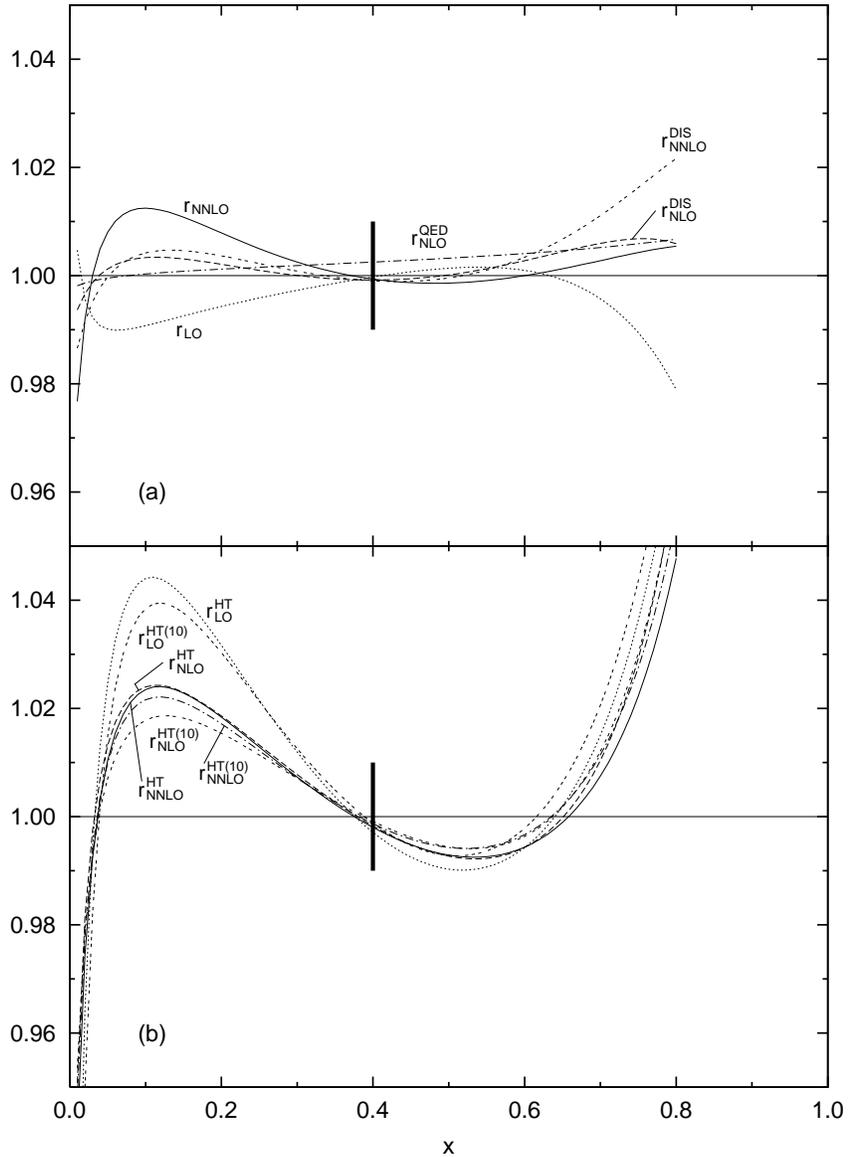}
\caption{The size of perturbative (a) and nonperturbative (b) 
uncertainties encountered in various perturbative $\alpha_s$--orders
of QCD analyses relative to our nominal NLO analysis 
of $F_{2,\mathrm{NLO}}^{ep}(x,Q^2)$ which always appears in the 
denominator of the ratios $r$ as defined and discussed in the text. 
The HT contributions to the fits shown in (b) refer to the cut
$Q^2\geq 4$ GeV$^2$, whereas HT(10) refers to a cut $Q^2\geq 10$
GeV$^2$.
The typical relative experimental accuracy is illustrated at $x=0.4$ 
by the vertical bar ($\pm1\,\%$). At larger and smaller values of 
$x$ the experimental error increases (e.g., the uncertainty is about 
$\pm2.5\,\%$ at $x=0.55$, and $\pm10\,\%$ at $x=0.18$). All results 
are shown for $Q^2=40\,\mathrm{GeV}^2$, but the agreement with 
data at $Q^2=4\,\mathrm{GeV}^2$ and $Q^2=100\,\mathrm{GeV}^2$ is 
similar.}
\end{figure}
\end{document}